\documentclass[12pt]{article}

\usepackage{epsfig}
\usepackage{cite}
\usepackage{amsmath, amssymb, amsfonts}
\usepackage{color}
\usepackage{latexsym}
\usepackage{graphicx}
\usepackage[colorlinks,bookmarks]{hyperref}
\hypersetup{pdfpagemode=UseNone, pdfstartview=FitH, linkcolor=blue,
            citecolor=red, urlcolor=blue}

\def\be{\begin{equation}}
\def\ee{\end{equation}}
\def\ba{\begin{eqnarray}}
\def\ea{\end{eqnarray}}

\def\bdm{\begin{displaymath}}
\def\edm{\end{displaymath}}
\def\la{~\mbox{\raisebox{-.6ex}{$\stackrel{<}{\sim}$}}~}

\def\bq{\begin{quote}}
\def\eq{\end{quote}}

 at 10truept
 at 14truept

\newcommand{\p}{\partial}

\newcommand{\Mpl}{M_{\mathrm{Pl}}}
\newcommand{\mps}{M_{\mathrm{Pl}}^2}

\newcommand{\bea}{\begin{eqnarray}}
\newcommand{\eea}{\end{eqnarray}}

\newcommand{\bi}{\begin{itemize}}
\newcommand{\ei}{\end{itemize}}

\newcommand{\beq}{\begin{equation}}
\newcommand{\eeq}{\end{equation}}
\newcommand{\beqa}{\begin{eqnarray}}
\newcommand{\eeqa}{\end{eqnarray}}
\newcommand{\mpl}{\Mpl}

\def\la{~\mbox{\raisebox{-.6ex}{$\stackrel{<}{\sim}$}}~}

\def\ltap{\ \raise.3ex\hbox{$<$\kern-.75em\lower1ex\hbox{$\sim$}}\ }
\def\gtap{\ \raise.3ex\hbox{$>$\kern-.75em\lower1ex\hbox{$\sim$}}\ }
\def\gl{\ \raise.5ex\hbox{$>$}\kern-.8em\lower.5ex\hbox{$<$}\ }
\def\roughly#1{\raise.3ex\hbox{$#1$\kern-.75em\lower1ex\hbox{$\sim$}}}

\begin{document}

\thispagestyle{empty}
\begin{flushright}
February 2022 \\
\end{flushright}
\vspace*{1.5cm}
\begin{center}

{\Large \bf  Hidden Variables of Gravity and Geometry and}
\vskip.3cm
{\Large \bf the Cosmological Constant Problem}  
\vskip.3cm

\vspace*{1.15cm} {\large 
Nemanja Kaloper$^{a, }$\footnote{\tt
kaloper@physics.ucdavis.edu} 
}\\
\vspace{.5cm}
{\em $^a$QMAP, Department of Physics and Astronomy, University of
California}\\
\vspace{.05cm}
{\em Davis, CA 95616, USA}\\

\vspace{1.65cm} ABSTRACT
\end{center}
We extend General Relativity by 
promoting Planck scale and the cosmological constant into
integration constants, interpreted as fluxes of $4$-forms hiding in the theory. When we include 
the charges of the $4$-forms, these `constants' can vary discretely from region to region. 
We explain how the cosmological constant problem can be solved in this new framework. 
When the cosmological constant picks up contributions from two different $4$-forms, with an irrational 
ratio of charges, the spectrum of its values is a very fine discretuum. When the charges are mutually irrational, 
$\frac{2\kappa_{\tt eff}^2 \kappa^2 |{\cal Q}_i|}{3{\cal T}^2_i} < 1$,  the discharge processes populating our 
discretuum will dynamically relax $\Lambda$, ceasing as $\Lambda$ approaches zero. 
Thus the theory exponentially favors a huge hierarchy  $\Lambda/\mpl^4 \ll 1$ instead of 
$\Lambda/\mpl^4 \simeq 1$. 

\vfill \setcounter{page}{0} \setcounter{footnote}{0}

\vspace{1cm}
\newpage

\vspace{1cm}

In this Letter, we discuss the salient features of our recent generalization \cite{Kaloper:2022utc} 
of standard General Relativity (GR) \cite{Hilbert:1915tx,Einstein:1915ca} where 
Planck scale, the cosmological term, and the matter sector couplings 
are discretely varying quantities throughout spacetime. 
We note that Hilbert's formulation of GR \cite{Hilbert:1915tx} can be 
extended using an arbitrary $4$-form ${\cal F} = d{\cal A}$ as the covariant integration measure in 
the action, which remains locally a constant multiple of the standard 
measure $\sqrt{g} d^4x$ when we add another $4$-form ${\cal G}=d{\cal B}$, coupled to ${\cal F}$ via the interaction   
$ {\cal F} \, \frac{\epsilon^{\mu\nu\lambda\sigma}}{\sqrt{g}} {\cal G}_{\mu\nu\lambda\sigma}$. 
This ensures that the ratio ${\cal F}/d^4x \sqrt{g}$ is constant, being the local value of Planck scale. The 
flux of ${\cal G}$ adds to the cosmological constant. Our Pancosmic GR 
generalizes\footnote{Some alternatives to minimal measure in the action were noted in 
\cite{Guendelman:1996qy,Gronwald:1997ei,Wilczek:1998ea,Kaloper:2015jra,DAmico:2017ngr,Kaloper:2018kma,Benisty:2019znu,Lee:2019efp,Cribiori:2020wch}.} 
the so-called ``unimodular" formulation of GR
\cite{Einstein:1919gv,Anderson:1971pn,Aurilia:1980xj,Duff:1980qv,Buchmuller:1988wx,Buchmuller:1988yn,Henneaux:1989zc,Ng:1990xz,Fiol:2008vk}. 
Finally we couple matter to geometry using 
the conformal $4$-form/matter coupling, 
which is locally indistinguishable from GR, and is form-invariant in the QFT loop 
expansion \cite{Englert:1975wj,Arkani-Hamed:2000hpr}. The theory 
is safe from ghosts in this particular case \cite{Kaloper:2022utc}. This provides at
least the first step in the demonstration that our mechanism may have a unitary UV
completion into some more fundamental approach to quantizing gravity.

When we include membranes charged under $4$-forms ${\cal F}$ and ${\cal G}$
their quantum-mechanical nucleation lead to local changes 
of the fluxes of ${\cal F}$ and ${\cal G}$, which jump up or down in the regions  
inside the membranes. As a result in the interior of these membrane-bound bubbles the effective strength 
of gravity and the value of cosmological 
constant, and also the values of couplings and scales of local matter theory jump relative to the outside. 

A sequence of such nested expanding bubbles will scan both $\kappa^2$ and $\Lambda$ over a wide  
range of parameters, realizing a toy model of the multiverse of eternal inflation \cite{Linde:2015edk}. This 
toy multiverse can solve the cosmological
constant problem \cite{Zeldovich:1967gd,Wilczek:1983as,Weinberg:1987dv}. To this end 
we include another $4$-form which is degenerate with 
the cosmological constant, with its corresponding system of 
charges. When the ratio of charges of our two $4$-forms which add up in the total cosmological 
constant is an irrational number the spectrum of values of $\Lambda$ is a very fine discretuum \cite{Banks:1991mb}. 
When $\frac{2\kappa_{\tt eff}^2 \kappa^2 |{\cal Q}_i|}{3{\cal T}^2_i} < 1$, which is the robust parameter ratio choice, 
the bounce action which controls the rates of the transitions that change $\Lambda$ has a pole at $\Lambda = 0$, driving
$\Lambda$ to $0^+$ dynamically, and stopping there. 

In this case, the distribution of the terminal values is effectively described by the semiclassical 
Euclidean partition function \cite{Hawking:1981gd,Baum:1983iwr,Hawking:1984hk,Abbott:1984qf}, which has an essential 
singularity\footnote{This requires a careful evaluation of the Euclidean action \cite{Duff:1989ah,Duncan:1989ug}.} 
at $\Lambda/\kappa^4_{\tt eff} \rightarrow 0$. The theory exponentially favors vacua with 
$\frac{\Lambda_{total}}{\mpl^4} \rightarrow 0 \ll 1$. The Newton's constant can be fixed separately, by observation, or by
using a `weak' anthropic principle. Cosmological constant, however, is driven to zero without any need for
anthropics. The theory avoids the empty universe problem of
\cite{Abbott:1984qf}, and can admit inflation. The cosmological 
constant problem reduces to ``{\it Why now?"}, whose answers 
might involve late time physics. 

Our choice of the bilinear dependence of the theory on the flux variables, as opposed to other powers, might seem 
special, even fine tuned. That is not so. Even with higher order corrections, which are weighted by $\Mpl$, the bilinear 
terms remain dominant for sub-Planckian fluxes and the same behavior as in the purely bilinear case
remains. In fact the higher-order corrections could come in with different coefficients for the two flux sectors, inducing 
mutually irrational variation of fluxes even if the actual ratio of charges were rational. For simplicity, therefore,
we will only work with linear dependence here. 

We note that our mechanism also avoids naturally the venerated Weinberg's 
no-go theorem \cite{Weinberg:1987dv} for the adjustment of the cosmological
constant, by exploiting loopholes in the assumption of the theorem. The adjustment occurs by quantum Brownian drift,
instead of smooth field variation. Further the evolution involves a special point in phase space, $\Lambda = 0^+$, which is the
quantum attractor where the bubble nucleation stops. These two features alone invalidate the theorem of \cite{Weinberg:1987dv}. 

A simplified example of the action which generalizes Einstein-Hilbert action \cite{Hilbert:1915tx,Einstein:1915ca} is \cite{Kaloper:2022utc}
\be
S = \int {\cal F} \Bigl( R - \frac{1}{4!} \frac{\epsilon^{\mu\nu\lambda\sigma}}{\sqrt{g}} {\cal G}_{\mu\nu\lambda\sigma} \Bigr) 
- \int d^4x \sqrt{g} \, {\cal L}_{\tt QFT} + S_{\tt membranes}\, .
\label{actionnew}
\ee
Here ${\cal F} = d{\cal A}$, ${\cal G}= d{\cal B}$, and ${\cal M}$ is a UV scale controlling the maximal flux the theory can admit. 
Careful variation yields field equations, which without membrane sources are
\be \frac{\epsilon^{\rho\zeta\gamma\delta}}{4!\sqrt{g}} {\cal F}_{\rho\zeta\gamma\delta}  \Bigl(
\frac{{\epsilon^{\alpha\beta\lambda\sigma}}{\cal G}_{\alpha\beta\lambda\sigma}}{ 4! \sqrt{g}} 
\delta^\mu{}_\nu -2R^\mu{}_\nu \Bigr) = T^\mu{}_\nu \, , ~~  
R - \frac{\epsilon^{\mu\nu\lambda\sigma}}{4!\sqrt{g}} {\cal G}_{\mu\nu\lambda\sigma} = 2 \lambda \, , ~~
- \frac{\epsilon^{\mu\nu\lambda\sigma}}{4!\sqrt{g}} {\cal F}_{\mu\nu\lambda\sigma} = \frac{\kappa^2}{2} \, . 
\label{alleqs}
\ee
The last two equations are integrated variations with respect to ${\cal A}$ and ${\cal B}$, and include the integration 
constants $\kappa^2$ and $\lambda$. A direct substitution of these two into the first gives
\be
\kappa^2 \Bigl(R^\mu{}_\nu - \frac12 R \, 
\delta^\mu{}_\nu \Bigr) = - \kappa^2 \lambda \, \delta^\mu{}_\nu + T^\mu{}_\nu \, ,
\label{tenseqsgr}
\ee
which are the local equations of GR. Yet both Planck scale and the
cosmological constant are integration constants.  Our equations (\ref{alleqs}), (\ref{tenseqsgr}) 
describe an infinity of GRs parameterized by $\kappa^2$, $\lambda$, which do not mix
with each other until the membranes are turned on. 

The membrane action is 
\be
S_{\tt membranes} =- {\cal T}_A \int d^3 \xi \sqrt{\gamma}_A - {\cal Q}_A \int {\cal A}  
- {\cal T}_B \int d^3 \xi \sqrt{\gamma}_B - {\cal Q}_B \int {\cal B}  \, .
\label{actionnewmem}
\ee
where 
\be
\int {\cal A} = \frac16 \int d^3 \xi {\cal A}_{\mu\nu\lambda} \frac{\p x^\mu}{\p \xi^\alpha} \frac{\p x^\nu}{\p \xi^\beta} 
\frac{\p x^\lambda}{\p \xi^\gamma} \epsilon^{\alpha\beta\gamma} \, ,
\ee
and likewise for ${\cal B}$. Here ${\cal T}_i, {\cal Q}_i$ are the membrane tension and charge, while 
$\xi^\alpha$ are the intrinsic membrane coordinates and 
embedding maps  are $x^\mu = x^\mu(\xi^\alpha)$. We take ${\cal T}_i >0$ to exclude negative local energy. 

Classically, membranes are sources fixed by initial conditions. However, quantum mechanically 
they can nucleate in background fields \cite{Brown:1987dd,Brown:1988kg}. 
This changes the distribution of sources and the evolution of bubble interiors. 
As a result any background will evolve by growing subspaces with different $\kappa^2$ and $\lambda$. 

The analysis simplifies when we transition to the 
magnetic dual variables of the $4$-forms ${\cal F}$ and ${\cal G}$ \cite{Kaloper:2022utc}. 
We use Lagrange multipliers
and rewrite the $4$-form sector of (\ref{actionnewmem})
in the first order formalism. Each pair of variables  
${\cal F}$,  ${\cal A}$ and ${\cal G}$, ${\cal B}$ are treated as independent dynamical variables and 
${\cal F} = d{\cal A}$ and ${\cal G} = d{\cal B}$ follow as constraints enforced by Lagrange multipliers, 
${\cal P}_A, {\cal P}_B$. The general first order path integral is
\be
Z = \int \ldots [{\cal D} {\cal A}] [{\cal D} {\cal B}] [{\cal D}{\cal F}] [{\cal D} {\cal G}] [{\cal D}{\cal P}_A] 
[{\cal D} {\cal P}_B] \, e^{i S({\cal A}, {\cal B}, {\cal F}, {\cal G}, ...) + i\int  {\cal P}_A ( {\cal F} -d {\cal A})
+ i\int  {\cal P}_B ( {\cal G} -d {\cal B})} \ldots \, .
\label{partf} 
\ee
Different dual pictures ensue from different order of integration, as in e.g. the formulations of flux monodromy models 
of inflation \cite{ks1,ks2,ks3}. So to dualize we integrate out ${\cal F}$ and ${\cal G}$, and redefine the Lagrange multipliers 
according to ${\cal P}_A = 2 \lambda$, ${\cal P}_B = \frac{\kappa^2}{2}$. 
In terms of the dual variables our action, with the membrane terms from (\ref{actionnewmem}), is \cite{Kaloper:2022utc}
\ba
S &=& \int d^4x \Bigl\{\sqrt{g} \Bigl(\frac{\kappa^2}{2} R - \kappa^2 \lambda 
- {\cal L}_{\tt QFT} \Bigr)- \frac{\lambda}{3} {\epsilon^{\mu\nu\lambda\sigma}} \partial_\mu {\cal A}_{\nu\lambda\sigma}
- \frac{\kappa^2}{12} {\epsilon^{\mu\nu\lambda\sigma}}\partial_\mu {\cal B}_{\nu\lambda\sigma} \Bigr\}  \nonumber \\
&& + \,\, S_{\tt boundary} - {\cal T}_A \int d^3 \xi \sqrt{\gamma}_A - {\cal Q}_A \int {\cal A}  
- {\cal T}_B \int d^3 \xi \sqrt{\gamma}_B - {\cal Q}_B \int {\cal B}  \, .
\label{actionnewmemd} 
\ea
$S_{\tt boundary}$ is a generalization of the 
Israel-Gibbons-Hawking boundary action \cite{israel,gibbhawk,gibbhawkcosm}, 
\be
S_{\tt boundary} = \int d^3 \xi \Bigr( [\frac{\lambda}{3} {\epsilon^{\alpha\beta\gamma}} {\cal A}_{\alpha\beta\gamma}]
+ [\frac{\kappa^2}{12} {\epsilon^{\alpha\beta\gamma}} {\cal B}_{\alpha\beta\gamma}] \Bigr)  - \int d^3 \xi \sqrt{\gamma} [\kappa^2 K ]\, .
\label{boundact}
\ee
$[...]$ is the jump across a membrane; $\lambda, \kappa^2$ are inside of $[...]$ since they jump if a ${\cal Q}_i$ is emitted. 

To include the matter sector, we replace the matter Lagrangian of our Eq. (\ref{actionnew}) by 
$\sqrt{g} {\cal L}_{\tt QFT}(g^{\mu\nu}) \rightarrow \sqrt{\hat g} {\cal L}_{\tt QFT}(\hat g^{\mu\nu})$, 
where $\hat g_{\mu\nu} = g_{\mu\nu} \sqrt{\frac{\kappa}{{\cal M}}}$  and, as noted, ${\cal M}$ 
is a UV scale controlling the perturbative expansion of the full effective action in the powers of
${\cal F}$. This ensures the theory retains a positive definite gravitational coupling which keeps ghosts at bay \cite{Kaloper:2022utc}. 
The matter loop corrections preserve this form of the action, as long as 
the regulator depends on $\kappa/{\cal M}$ in the same way as the matter Lagrangian 
\cite{Englert:1975wj,Arkani-Hamed:2000hpr}. 
We can also generalize the gravitational sector, substituting 
$\frac{\kappa^2}{2} R \rightarrow \frac{\mps + \kappa^2}{2} R \,$:
\ba
S &=& \int \Bigl\{\sqrt{g} \Bigl(\frac{\mps+\kappa^2}{2} R-\kappa^2 \lambda 
-\frac{\kappa^2}{{\cal M}^2} {\cal L}_{\tt QFT}(\frac{{\cal M}}{\kappa}{g^{\mu\nu}})\Bigr)
-\frac{\lambda}{3} {\epsilon^{\mu\nu\lambda\sigma}} \partial_\mu {\cal A}_{\nu\lambda\sigma}
- \frac{\kappa^2}{12} {\epsilon^{\mu\nu\lambda\sigma}}\partial_\mu {\cal B}_{\nu\lambda\sigma} \Bigr\}  \nonumber \\
&&~~~~ + \,\, S_{\tt boundary} - {\cal T}_A \int d^3 \xi \sqrt{\gamma}_A - {\cal Q}_A \int {\cal A}  
- {\cal T}_B \int d^3 \xi \sqrt{\gamma}_B - {\cal Q}_B \int {\cal B}  \, . 
\label{actionnewmemdconf}
\ea
The variation of (\ref{actionnewmemdconf}) yields \cite{Kaloper:2022utc}
\ba
&&~~~~~~~~~~ (\mps + \kappa^2) G^\mu{}_\nu =  - \kappa^2 \lambda \, \delta^\mu{}_\nu + T^\mu{}_\nu + \ldots\, , ~~~~~~~  
\hat {\cal F}_{\mu\nu\lambda\sigma} = \frac{\kappa^2}{2} \sqrt{g} \, {\epsilon_{\mu\nu\lambda\sigma}} \, ,   \\
&& 
\hat {\cal G}_{\mu\nu\lambda\sigma} = \frac{2 \kappa^2\lambda- \kappa^2 R - T/4}{4\kappa^2} 
\sqrt{g} \, {\epsilon_{\mu\nu\lambda\sigma}} \, , ~ 2 n^\mu \partial_\mu \lambda  = {\cal Q}_A \delta(r-r_0) \, , ~ 
\frac12 n^\mu \partial_\mu \kappa^2 = {\cal Q}_B \delta(r-r_0) \, .  \nonumber 
\label{tenseqsgrmagdualconf}
\ea
The Israel-Gibbons-Hawking action changes slightly by $\kappa^2 \rightarrow \kappa^2_{\tt eff} = \mps + \kappa^2$.

To Wick-rotate the action we use $t = - i x^0_E$, and turn the crank \cite{Kaloper:2022utc}. 
Defining the Euclidean action by $i S = - S_E$ and restricting to locally maximally symmetric backgrounds, 
$\langle {\cal L}^E_{\tt QFT} \rangle = \Lambda_{\tt QFT}$, 
with $\Lambda_{\tt QFT}$ the matter sector vacuum energy to an arbitrary loop order,
\ba
S_E&=&\int d^4x_E \Bigl\{\sqrt{g} \Bigl(-\frac{\kappa^2_{\tt eff}}{2} R_E + \kappa^2 \lambda 
+ \Lambda_{\tt QFT} \Bigr)- \frac{\lambda}{3} {\epsilon^{\mu\nu\lambda\sigma}_E} \partial_\mu {\cal A}^E_{\nu\lambda\sigma}
- \frac{\kappa^2}{12} {\epsilon^{\mu\nu\lambda\sigma}_E}\partial_\mu {\cal B}^E_{\nu\lambda\sigma} \Bigr\} \nonumber \\
\label{actionnewmemeu}
&& +~S_{\tt boundary} + {\cal T}_A \int d^3 \xi_E \sqrt{\gamma}_A - \frac{{\cal Q}_A}{6} \int d^3 \xi_E \, {\cal A}^E_{\mu\nu\lambda} \, 
\frac{\p x^\mu}{\p \xi^\alpha} \frac{\p x^\nu}{\p \xi^\beta} 
\frac{\p x^\lambda}{\p \xi^\gamma} \epsilon_E^{\alpha\beta\gamma} \\
&& ~~~~~~+~  {\cal T}_B \int d^3 \xi_E \sqrt{\gamma}_B  - \frac{{\cal Q}_B}{6} \int d^3 \xi_E \, 
{\cal B}^E_{\mu\nu\lambda} \, \frac{\p x^\mu}{\p \xi^\alpha} \frac{\p x^\nu}{\p \xi^\beta} 
\frac{\p x^\lambda}{\p \xi^\gamma} \epsilon_E^{\alpha\beta\gamma} \, . \nonumber
\ea
With our couplings, 
$\Lambda_{\tt QFT} = \kappa^2 \frac{{\cal M}_{\tt UV}^4}{{\cal M}^2} + \ldots = \kappa^2 {\cal H}^2_{\tt QFT}$, where
${\cal M}_{UV}^4$ is the QFT UV cutoff and ellipsis denote subleading terms \cite{Englert:1975wj,Arkani-Hamed:2000hpr}. 
Thus $\Lambda = \kappa^2 \bigl(\lambda + {\cal H}^2_{\tt QFT}\bigr) = \kappa^2 \lambda_{\tt eff}$.

The membrane-induced transitions link geometries with $\kappa^2_{out/in}, \Lambda_{out/in}$, 
({\it out/in} denote parent and offspring geometries (exterior and interior of 
the membranes, respectively)). The configurations with local $O(4)$ symmetry
dominate since they have minimal Euclidean action \cite{Coleman:1977py,Callan:1977pt,Coleman:1980aw}. 
Both {\it in} and {\it out} can be described with the metrics $ds^2_E =  dr^2 + a^2(r) \, d\Omega_3$ 
where $d\Omega_3$ is the line element on a unit $S^3$. Here $a$ solves $\bigl(\frac{a'}{a}\bigr)^2 - \frac{1}{a^2} 
= - \Lambda/3 \kappa^2_{\tt eff}$. The prime is the $r$-derivative \cite{Kaloper:2022utc}.
The variations of (\ref{actionnewmemeu}) give also the membrane-induced contributions. 
We find \cite{Kaloper:2022utc} 
\be
\lambda_{out} - \lambda_{in}  = \frac{{\cal Q}_A}{2} \, , ~~~~ \kappa^2_{out} - \kappa_{in}^2 = 2 {\cal Q}_B \, ,
~~~~ \kappa^2_{{\tt eff}~{out}} \frac{a_{out}'}{a} - \kappa^2_{{\tt eff}~{in}} \frac{a_{in}'}{a} 
 = -\frac{{\cal T}_A + {\cal T}_B}{2} \, .
 \label{metricjc}
\ee 
We wrote the discontinuities as if the emission of $A$ and $B$ membranes were co-located simply to save space.
We ignore the discontinuity in ${\cal B}$ since the $p$-form terms in the action
cancel out exactly against the charges and do not affect the bounce action \cite{Kaloper:2022utc}. 

In semiclassical approximation, the membrane nucleation rates are 
$\Gamma \sim e^{-S_{bounce}}$ \cite{Coleman:1977py,Callan:1977pt,Coleman:1980aw},
where $S({\tt bounce}) = S({\tt instanton}) - S({\tt parent})$. Instanton actions are the Euclidean actions computed on a solution
with a number of membranes, starting with one and counting off. Their classification was initiated in 
\cite{Brown:1987dd,Brown:1988kg} for the theories with $4$-form fluxes screening 
the cosmological constant \cite{Brown:1987dd,Brown:1988kg,Duncan:1989ug,Bousso:2000xa,Feng:2000if}. 
We find important differences in our case because the cosmological constant depends 
on membrane charges linearly, instead of quadratically \cite{Kaloper:2022utc}. 

\begin{figure}[bth]
    \centering
    \includegraphics[width=7.5cm]{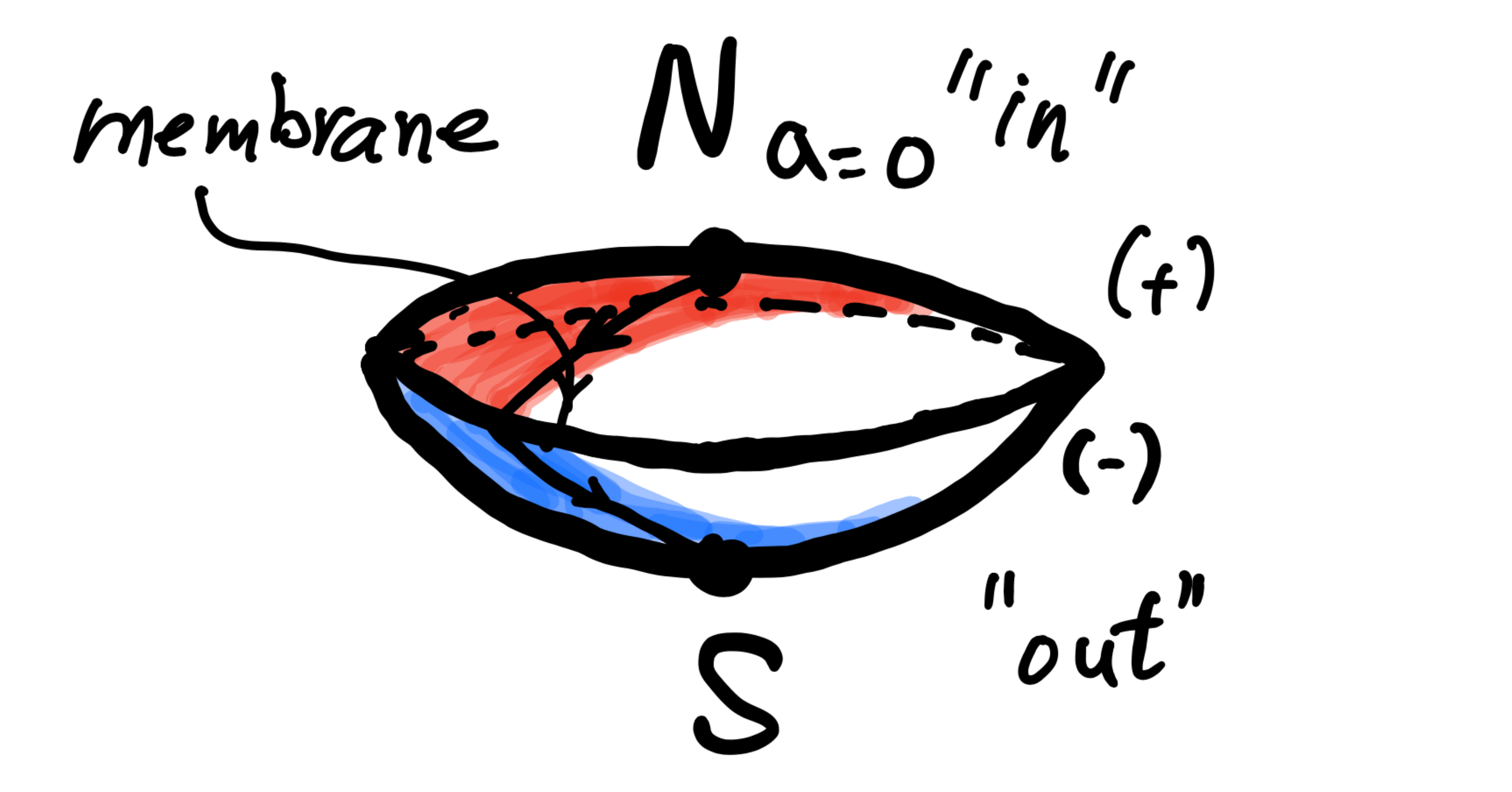}
    \caption{A $q<1$ instanton comprised of two sections of $S^4$.}
    \label{fig1}
\end{figure}
We focus 
here on the most interesting case of $dS \rightarrow dS$ decay 
when $q = \frac{2\kappa_{\tt eff}^2 \kappa^2 |{\cal Q}_A|}{3{\cal T}^2_A} < 1$. In this case
only the $(-+)$ instantons of Fig. (\ref{fig1}) are kinematically allowed for $A$-membrane 
transitions \cite{Kaloper:2022utc}, and the geometries can be glued together to form an instanton 
if\footnote{Corresponding to $\zeta_{out} = -1$, $\zeta_{in} = +1$ in 
the notation of \cite{Brown:1987dd,Brown:1988kg,Kaloper:2022utc}.}
\be
\sqrt{ 1-  \frac{\Lambda_{out} a^2}{3 \kappa_{\tt eff}^2}} 
= \frac{{\cal T}_A}{4\kappa_{\tt eff}^2}\Bigl(1 -  q \Bigr)\, a \, , ~~~~~~~~ 
\sqrt{1-  \frac{\Lambda_{in} a^2}{3 \kappa_{\tt eff}^2}} 
= \frac{{\cal T}_A }{4\kappa_{\tt eff}^2}\Bigl(1 +q \Bigr) \, a \, . 
\label{diffroots}
\ee
\underbar{\it Crucially}, due to (bi)linear dependence on charges -- instead of quadratic -- 
other $dS \rightarrow dS$ instantons are \underbar{\it forbidden} 
when $q<1$ \cite{Kaloper:2022utc}. This has dramatic consequences: as a result, the resulting theory has a quantum attractor
at $\Lambda = 0^+$. 

We now uncover this behavior. First, it turns out to be useful to solve Eqs.
(\ref{diffroots}) for $a^2$, which is the radius of the membrane connecting parent and offspring. We find 
\be
\frac{1}{a^2} = \frac{\Lambda_{out}}{3\kappa_{\tt eff}^2} + \Bigl(\frac{{\cal T}_A}{4 \kappa_{\tt eff}^2}\Bigr)^2 
\Bigr(1 -q\Bigr)^2 
= \frac{\Lambda_{in}}{3\kappa_{\tt eff}^2} + \Bigl(\frac{{\cal T}_A}{4 \kappa_{\tt eff}^2}\Bigr)^2 
\Bigr(1 + q \Bigr)^2 \, .
\label{radii}
\ee
Hence there are two regimes of bubble formation. When $a^2$ is comparable to 
de Sitter radii, Eq. (\ref{radii}) 
shows that $\sim (1-\frac{\Lambda_j a^2}{3 \kappa^2_{\tt eff}})^{1/2} \ll 1$
and the bounce action is approximated by the difference of one half of the
parent and offspring  horizon areas divided by $4G_N$, 
\be 
S_{\tt bounce} \simeq - \frac{12\pi^2 \kappa^4_{\tt eff} \Delta \Lambda}{\Lambda_{out} \Lambda_{in}} \, , ~~~~~~~~ 
\Delta \Lambda = \Lambda_{out} - \Lambda_{in} = \frac12 \kappa^2  {\cal Q}_A \, .
\label{fastbounce}
\ee
Since $q<1$, as long as 
$\Lambda_{out} \gg 3 \kappa^2_{\tt eff} \Bigl(\frac{{\cal T}_A}{4 \kappa_{\tt eff}^2}\Bigr)^2$, 
in this regime the discharge of the cosmological constant is fast because $S_{\tt bounce} <0$. 
The cosmological constant is repelled down from the Planckian scales. 
The reverse processes increasing $\Lambda$ ($\Delta \Lambda < 0$) 
also occur but their bounce action is the negative of (\ref{fastbounce}) and they are 
rarer. The overall trend is the decrease of $\Lambda$. 

This stage ends when 
$\Lambda < 3 \kappa^2_{\tt eff} \Bigl(\frac{{\cal T}_A}{4 \kappa_{\tt eff}^2}\Bigr)^2$, 
after which nucleations proceed via production of small bubbles, with the 
bounce action \cite{Kaloper:2022utc}
\be
S_{\tt bounce}  \simeq \frac{24\pi^2 \kappa^4_{\tt eff}}{\Lambda_{out}} \Bigl(1
- \frac{8}{3} \frac{\kappa^2_{\tt eff} \Lambda_{out}}{ {\cal T}_A^2} \Bigr)\, ,
\label{familiars2}
\ee 
and $S_{\tt bounce} > 0$ because $\Lambda < 3 \kappa^2_{\tt eff} \Bigl(\frac{{\cal T}_A}{4 \kappa_{\tt eff}^2}\Bigr)^2$. 
Therefore $S_{\tt bounce}$ \underbar{\it 
diverges} as $\Lambda_{out} \rightarrow 0^+$ -- as we asserted above. So the bubbling rate $\Gamma \sim e^{-S_{bounce}}$
has an essential singularity at $\Lambda_{out} \rightarrow 0^+$, where the rate goes to zero. 
Hence when $q < 1$ small $\Lambda$ is 
metastable, and the locally Minkowski  space is 
absolutely stable to discharges. 

Once $q <1$, the only way it can be disturbed is if $\kappa^2_{\tt eff}$ increases by the emission of ${\cal Q}_B$. Another
possible problem is if emissions yield $\kappa^2_{\tt eff}<0$, leading to a pandemic of spin-2 ghosts.
Arranging the values of ${\cal Q}_B$ and ${\cal T}_B$ we can prevent both of these dangers. The 
dangerous processes are mediated by large bubbles, which are completely blocked when 
$16 \, \frac{\kappa^4_{{\tt eff} }|{\cal Q}_B|}{{\cal T}^2_B}  \ll 1 $. In this limit the $B$-bubble radius at nucleation is 
\be
\frac{1}{a^2} =  
\bigl(\frac{{\cal T}_B}{4\kappa^2_{{\tt eff}~out}} \bigr)^2 \Bigl(1+ {\cal O}\bigl( \frac{\kappa^4_{{\tt eff}~out}  {\cal Q}_B}{{\cal T}^2_B} \bigr) \Bigr) =
\bigl(\frac{{\cal T}_B}{4\kappa^2_{{\tt eff}~in}} \bigr)^2 \Bigl(1+ {\cal O}\bigl(\frac{\kappa^4_{{\tt eff}~in}  {\cal Q}_B}{{\cal T}^2_B}\bigr) \Bigr) \, . 
\label{radiib}
\ee
Since only small bubbles can nucleate when 
$ \frac{\Lambda}{3\kappa^4_{{\tt eff}}}  < \Bigl(\frac{{\cal T}_B}{4\kappa^3_{{\tt eff}}}\Bigr)^2$, 
$\kappa^2_{\tt eff}$ stays constant. 
The condition $16 \, \frac{\kappa^4_{{\tt eff} }|{\cal Q}_B|}{{\cal T}^2_B}  \ll 1 $ also blocks the
bubble size from below, ensuring that $a > 1/\kappa_{\tt eff}$. This regime is disconnected both from
Planckian physics and from $\kappa^2<0$, at least in the 
semiclassical approximation which we pursue \cite{Kaloper:2022utc}. 
Further in this limit kinematics restricts $dS \rightarrow dS$ 
discharges to only be mediated by instantons of Fig. (\ref{fig1}). Hence ${\cal Q}_B$ emissions 
again completely shut off at Minkowski. Finally, by arranging 
$\frac{\kappa^4_{{\tt eff}~in}  {\cal Q}_B}{{\cal T}^2_B} \ll \frac{\kappa^4_{{\tt eff}~in}  {\cal Q}_A}{{\cal T}^2_A} < 1$ 
we can make ${\cal Q}_B$ discharges a lot slower than ${\cal Q}_A$ discharges.

So in our generalization of GR both the cosmological constant and Newton's constant 
change by membrane emissions. The discharges are quantum-mechanical and nonperturbative, and cease in the classical limit and flat space,
$\Lambda/\kappa^4 \rightarrow 0$. This limit of the cosmological constant evolution is attained by quantum Brownian drift, and is the 
quantum attractor. This fits with ideas that an eternal, stable de Sitter space may not exist in a UV complete theory 
\cite{Banks:2000fe,Banks:2001yp,Witten:2001kn,Goheer:2002vf,Dvali:2017eba,Obied:2018sgi,Susskind:2021dfc}, and
reminds of the wormhole approach to quantum gravity 
 \cite{Hawking:1978pog,Coleman:1988tj,Coleman:1989ky,Giddings:1988wv,Fischler:1988ia,Fischler:1989ka,Banks:1984cw,Polchinski:1989ae,Horava:2000tb}. 

Let us now show how this discharges the cosmological constant. First we define the problem in our theory. The total cosmological 
constant is $\Lambda_{\tt total} = \kappa^2 \Bigl( \frac{{\cal M}_{\tt UV}^4}{{\cal M}^2} + \frac{V}{{\cal M}^2} + \lambda \Bigr)$,
where we include the QFT vacuum energy $\sim {\cal M}_{\tt UV}^4$, any nonvanishing QFT
potential $\sim V$, and our variable $\lambda$.
Since $\lambda$ and $\kappa^2$ change discretely, by $\Delta \lambda = {\cal Q}_A/2$, $\Delta \kappa^2 = 2{\cal Q}_B$, we have 
$\lambda = \lambda_0 + N \frac{{\cal Q}_A}{2}$, $\kappa^2 = \kappa^2_0 + 2 {\cal N} {\cal Q}_B$ with 
$N$ and ${\cal N}$ arbitrary integers. The net cosmological constant term is 
$\Lambda_{\tt total} = \bigl( \kappa^2_0 + 2 {\cal N} {\cal Q}_B \bigr)\bigl( \frac{\Lambda_0}{{\cal M}^2} + N \frac{{\cal Q}_A}{2} \bigr)$, 
where $\Lambda_0 = {\cal M}_{\tt UV}^4 + V + {\cal M}^2 \lambda_0$ and   
$\Delta \Lambda_{\tt total} = \bigl( \kappa^2_0 + 2 {\cal N} {\cal Q}_B \bigr){{\cal Q}_A}/{2}$. To get 
$\Lambda_{\tt total}/\kappa^4_{\tt eff} < 10^{-120}$, we must either pick a tiny ${\cal Q}_A$ or 
fine tune $\bigl(\kappa_0^2 + 2 {\cal N} {\cal Q}_B \bigr){\Lambda_0}/{{\cal M}^2}$. 
Variations of $\kappa^2_{\tt eff}$ cannot help since we can't suppress the curvature of the universe
without reducing the gravitational force between two masses. 

There is a simple solution to this obstacle, however. 
We add one more $4$-form which contributes to $\lambda$, but couple it 
to a different membrane with parameters ${\cal T}_{\hat A}, {\cal Q}_{\hat A}$:
\ba
{\cal S} = S - \int d^4x \sqrt{g} \Bigl(\kappa^2 \hat \lambda 
+ \frac{\hat \lambda}{3} {\epsilon^{\mu\nu\lambda\sigma}} \partial_\mu \hat {\cal A}_{\nu\lambda\sigma} \Bigr) 
- {\cal T}_{\hat A} \int d^3 \xi \sqrt{\gamma}_{\hat A} - {\cal Q}_{\hat A} \int \hat {\cal A}  \, . 
\label{actionnewext}
\ea
The membranes $\hat A$ behave exactly as the $A$ ones. We demand $\hat q < 1$ in addition to $q<1$, and that 
$\frac{{\cal Q}_{\hat A}}{{\cal Q}_A} = \omega$ is an irrational number, as in 
the irrational axion proposal \cite{Banks:1991mb} (see also \cite{Kaloper:2018kma}). The cosmological constant `quantization' law now becomes
\be
\Lambda_{\tt total} = \bigl( \kappa_0^2 + 2 {\cal N} {\cal Q}_B \bigr)\Bigl( \frac{\Lambda_0}{{\cal M}^2} + \frac{{\cal Q}_A}{2} \bigl( N + 
\hat N \omega \bigr) \Bigr) \, . 
\ee
Since $\omega$ is irrational, for any real number $\rho$ there exist integers $N, \hat N$
such that $N + \hat N \omega$ is arbitrarily close to $\rho$ \cite{Banks:1991mb,niven}. Thus 
integers $N, \hat N$ exist such that $N + \hat N \omega$ is arbitrarily close to $- \frac{2\Lambda_0}{{\cal Q}_A {\cal M}^2}$.
The set of $\Lambda_{total}$ is dense, with values arbitrarily close to zero! Further,
there is no axion `gauging' these dense discrete shifts and no emerging global symmetries \cite{Banks:1991mb,banksseiberg}. 
For any initial value of $\Lambda$, there exist many sequences of discharging membranes, in any order, 
which will yield terminal values $N, \hat N$ for which the cosmological constant is arbitrarily close to zero, and very long lived. 

Due to the irrational ratio of charges, {\it all} previously separated superselection sectors now mix together, 
transitioning between each other by utilizing both $A, \hat A$ charges. The slower nucleation processes when 
$\Lambda$ is well below the cutoff also allow up-jumps, which raise $\Lambda$, 
and so the superselection sectors will form a 
very fine discretuum. The states with $\mpl^4 \gg \Lambda >0$ will be very long lived, decaying  
predominantly to $\Lambda \rightarrow 0^+$. The key ingredient here is the pole of the bounce action, Eq. (\ref{familiars2}),
specific to the $(-+)$ of Fig. (\ref{fig1}). Its presence means that $\Lambda \rightarrow 0^+$ is the dynamical attractor. 
This is captured by the behavior of the semi-classical Euclidean partition function. 

Indeed, consider
$Z = \int \ldots {\cal DA} {\cal D} \hat {\cal A}  {\cal DB} {\cal D} \lambda {\cal D} \hat \lambda {\cal D} \kappa^2 {\cal D} g  \,  e^{-{\cal S}_E}$, 
the Euclidean variant of the magnetic dual partition function. In the saddle point approximation, 
\be
Z = \sum_{instantons} \sum_{\lambda, \hat \lambda, \kappa^2}  e^{-{\cal S}_E(instanton)} \, ,
\label{sumZ}
\ee
the idea was to sum over classical extrema of the action. In our case this 
begins with summing over the Euclidean instantons with any number of membranes included.
The $O(4)$ invariant solutions should minimize the action, and so this is a reasonable leading order approximation 
 \cite{Coleman:1977py,Callan:1977pt,Coleman:1980aw}. Thus $Z$ will invariably be dominated by our instantons. 

This sum is challenging, but we can get a feel for the individual terms. Inverting the 
bounce action, ${\cal S}({\tt instanton}) = {\cal S}({\tt bounce}) + {\cal S}({\tt parent})$.
Without offspring, the instanton action equals the parent action, which is just the negative of the horizon area 
divided by $4G_N$, ${\cal S}({\tt parent}) = - 24 \pi^2 \frac{\kappa^4_{\tt eff}}{\Lambda_{out}}$. For $n$ generations, 
summing over the family tree, ${\cal S}({\tt instanton},n) = \sum_n {\cal S}({\tt offspring}, n) + {\cal S}({\tt progenitor})$.
By ``offspring" we mean nested segments separated from the parent by membranes.
The ``progenitor" geometry is the primordial parent initiating the tree. 
Hence by Eq. (\ref{familiars2}) ${\cal S}({\tt instanton}) \la - 64 \pi^2 {\kappa_{\tt eff}^6}/{\cal T}_i^2$, and 
so an extended lineage may yield 
${\cal S}({\tt instanton},n) \rightarrow - 64 \pi^2 {\kappa_{\tt eff}^6} (\frac{n_A}{ {\cal T}_A^2} 
+ \frac{n_{\hat A}}{{\cal T}_{\hat A}^2} ) \rightarrow  - 24 \pi^2 \frac{\kappa^4_{\tt eff}}{\Lambda_{\tt terminal}}$.
The number of generations will be bounded by $- 24 \pi^2 \frac{\kappa^4_{\tt eff}}{\Lambda_{\tt terminal}}$, 
since nucleations can go on until $\Lambda_{\tt terminal} \rightarrow 0^+$, at which point the rate vanishes. 

Thus the sum (\ref{sumZ}), $Z \sim \sum e^{24 \pi^2 \frac{\kappa^4_{\tt eff}}{\Lambda} + \ldots}$, 
will be heavily skewed toward $\Lambda \rightarrow 0^+$, reflecting that discharges cease in the Minkowski limit. 
Ergo, in our generalization of GR, $Z$ heavily prefers 
$\frac{\Lambda}{\kappa^4_{\tt eff}} \rightarrow 0$; de Sitter is unstable, quantum mechanics + GR 
dynamically relax $\Lambda$ to zero. The 
instability stops as $\Lambda \rightarrow 0^+$, 
and the final (near) Minkowski space is extremely long lived.  
To explain 
$G_N = \frac{1}{8\pi \mps} \simeq 10^{-38} \, ({\rm GeV})^{-2}$ we could either just decide to measure it, or 
invoke a mild version of the 
`Weak Anthropic Principle', which can fix $G_N$ by placing the Earth in the habitable zone around the Sun.
 
The ``empty universe" problem \cite{Abbott:1984qf} is averted since the relaxation of $\Lambda$ 
involves large successive jumps, and the tiny terminal $\Lambda$ comes from
the irrational ratio of charges. So the cosmological constant does not always dominate, but just 
sometimes \cite{Bousso:2000xa}. The relaxation process is a random walk. 
Finally, we note that in addition to starting form a high scale, an empty universe could 
`restart' itself by a rare quantum jump which increases the cosmological 
constant  since the up-jumps are also possible\footnote{Note, that this
may have implications for the cosmological arrow of time, as it might yield a cosmic rebirth, rare as it is. As this issue is
irrelevant for the present discussion, we ignore it here, intending to return to it at a later time.}, and then in subsequent evolution an 
inflationary stage is found \cite{Garriga:2000cv}. In this sense, this channel for inflation may seem {\it \`a priori} rare, 
but since the system can continue exploring 
the phase space, even a `rare' event can be found eventually \cite{Carroll:2004pn}. 
In any case, our universe would evolve towards Minkowski, $\Lambda \rightarrow 0$, so 
that the potential problems with more likely, smaller scale fluctuations dubbed `Boltzmann Brains' may be 
avoided \cite{Carroll:2004pn,Bousso:2008hz,DeSimone:2008if,Susskind:2014rva}. 
Hence, the avoidance of the ``empty universe" problem opens the door for embedding `normal' cosmology in our
framework. We will return to the precise details of how to introduce inflation and postinflationary cosmology, 
as well as the questions about the UV completion of the mechanism in future work.

\vskip.3cm

{\bf Acknowledgments}: 
We would like to thank G. D'Amico and A. Westphal for valuable comments and discussions. 
NK is supported in part by the DOE Grant DE-SC0009999.


\begin{thebibliography}{99}

\bibitem{Kaloper:2022utc}
N.~Kaloper,
[arXiv:2202.08860 [hep-th]].

\bibitem{Hilbert:1915tx}
D.~Hilbert, 
Gott. Nachr. \textbf{27}, 395-407 (1915).

\bibitem{Einstein:1915ca}
A.~Einstein, 
Sitzungsber. Preuss. Akad. Wiss. Berlin 
\textbf{1915}, 844-847 (1915). 
\bibitem{Einstein:1919gv}
A.~Einstein,
Sitzungsber. Preuss. Akad. Wiss. Berlin (Math. Phys.) \textbf{1919}, 349-356 (1919). 

\bibitem{Anderson:1971pn}
J.~L.~Anderson and D.~Finkelstein,
Am. J. Phys. \textbf{39}, 901-904 (1971). 

\bibitem{Aurilia:1980xj}
A.~Aurilia, H.~Nicolai and P.~K.~Townsend,
Nucl. Phys. B \textbf{176}, 509-522 (1980). 

\bibitem{Duff:1980qv}
M.~J.~Duff and P.~van Nieuwenhuizen,
Phys. Lett. B \textbf{94}, 179-182 (1980). 

\bibitem{Buchmuller:1988wx}
W.~Buchmuller and N.~Dragon,
Phys. Lett. B \textbf{207}, 292-294 (1988). 

\bibitem{Buchmuller:1988yn}
W.~Buchmuller and N.~Dragon,
Phys. Lett. B \textbf{223}, 313-317 (1989).

\bibitem{Henneaux:1989zc}
M.~Henneaux and C.~Teitelboim,
Phys. Lett. B \textbf{222}, 195-199 (1989). 

\bibitem{Ng:1990xz}
Y.~J.~Ng and H.~van Dam,
J. Math. Phys. \textbf{32}, 1337-1340 (1991). 

\bibitem{Fiol:2008vk}
B.~Fiol and J.~Garriga,
JCAP \textbf{08}, 015 (2010)
[arXiv:0809.1371 [hep-th]].

\bibitem{Guendelman:1996qy}
E.~I.~Guendelman and A.~B.~Kaganovich,
Phys. Rev. D \textbf{53}, 7020-7025 (1996)
[arXiv:gr-qc/9605026 [gr-qc]].

\bibitem{Gronwald:1997ei}
F.~Gronwald, U.~Muench, A.~Macias and F.~W.~Hehl,
Phys. Rev. D \textbf{58}, 084021 (1998)
[arXiv:gr-qc/9712063 [gr-qc]].

\bibitem{Wilczek:1998ea}
F.~Wilczek,
Phys. Rev. Lett. \textbf{80}, 4851-4854 (1998)
[arXiv:hep-th/9801184 [hep-th]].

\bibitem{Kaloper:2015jra}
N.~Kaloper, A.~Padilla, D.~Stefanyszyn and G.~Zahariade,
Phys. Rev. Lett. \textbf{116}, no.5, 051302 (2016)
[arXiv:1505.01492 [hep-th]].

\bibitem{DAmico:2017ngr}
G.~D'Amico, N.~Kaloper, A.~Padilla, D.~Stefanyszyn, A.~Westphal and G.~Zahariade,
JHEP \textbf{09}, 074 (2017)
[arXiv:1705.08950 [hep-th]].

\bibitem{Kaloper:2018kma}
N.~Kaloper,
JHEP \textbf{11}, 106 (2019) 
[arXiv:1806.03308 [hep-th]].

\bibitem{Benisty:2019znu}
D.~Benisty, E.~Guendelman, A.~Kaganovich, E.~Nissimov and S.~Pacheva,
Springer Proc. Math. Stat. \textbf{335}, 239-252 (2019)
[arXiv:1905.09933 [gr-qc]].

\bibitem{Lee:2019efp}
H.~M.~Lee,
JHEP \textbf{01}, 045 (2020)
[arXiv:1908.04252 [hep-ph]].

\bibitem{Cribiori:2020wch}
N.~Cribiori, F.~Farakos and G.~Tringas,
JHEP \textbf{05}, 060 (2020)
[arXiv:2001.05757 [hep-th]].

\bibitem{Englert:1975wj}
F.~Englert, E.~Gunzig, C.~Truffin and P.~Windey,
Phys. Lett. B \textbf{57}, 73-77 (1975). 

\bibitem{Arkani-Hamed:2000hpr}
N.~Arkani-Hamed, S.~Dimopoulos, N.~Kaloper and R.~Sundrum,
Phys. Lett. B \textbf{480}, 193-199 (2000)
[arXiv:hep-th/0001197 [hep-th]].

\bibitem{Linde:2015edk}
A.~Linde,
Rept. Prog. Phys. \textbf{80}, no.2, 022001 (2017)
[arXiv:1512.01203 [hep-th]].

\bibitem{Zeldovich:1967gd}
Y.~B.~Zeldovich,
JETP Lett.\  {\bf 6}, 316 (1967);
Sov.\ Phys.\ Usp.\  {\bf 11}, 381 (1968).

\bibitem{Wilczek:1983as} 
F.~Wilczek,
Phys.\ Rept.\  {\bf 104}, 143 (1984).

\bibitem{Weinberg:1987dv} 
S.~Weinberg,
Rev.\ Mod.\ Phys.\  {\bf 61}, 1 (1989).

\bibitem{Banks:1991mb}
T.~Banks, M.~Dine and N.~Seiberg,
Phys. Lett. B \textbf{273}, 105-110 (1991)
[arXiv:hep-th/9109040 [hep-th]].

\bibitem{Hawking:1981gd}
S.~W.~Hawking,
``The Cosmological Constant and the Weak Anthropic Principle," 
Print-82-0177 (CAMBRIDGE). Contribution to
``Nuffield Workshop on Quantum Structure of Space and Time", 423-432 (1981).

\bibitem{Baum:1983iwr}
E.~Baum,
Phys. Lett. B \textbf{133}, 185-186 (1983). 

\bibitem{Hawking:1984hk}
S.~W.~Hawking,
Phys. Lett. B \textbf{134}, 403 (1984). 

\bibitem{Abbott:1984qf}
L.~F.~Abbott,
Phys. Lett. B \textbf{150}, 427-430 (1985). 

\bibitem{Duff:1989ah}
M.~J.~Duff,
Phys. Lett. B \textbf{226}, 36 (1989).

\bibitem{Duncan:1989ug}
M.~J.~Duncan and L.~G.~Jensen,
Nucl. Phys. B \textbf{336}, 100-114 (1990). 

\bibitem{Brown:1987dd}
J.~D.~Brown and C.~Teitelboim,
Phys. Lett. B \textbf{195}, 177-182 (1987). 

\bibitem{Brown:1988kg}
J.~D.~Brown and C.~Teitelboim,
Nucl. Phys. B \textbf{297}, 787-836 (1988). 

\bibitem{ks1}
N.~Kaloper and L.~Sorbo,
Phys. Rev. D \textbf{79}, 043528 (2009)
[arXiv:0810.5346 [hep-th]].

\bibitem{ks2}
N.~Kaloper and L.~Sorbo,
Phys. Rev. Lett. \textbf{102}, 121301 (2009)
[arXiv:0811.1989 [hep-th]].

\bibitem{ks3}
N.~Kaloper, A.~Lawrence and L.~Sorbo,
JCAP \textbf{03}, 023 (2011)
[arXiv:1101.0026 [hep-th]].

\bibitem{israel}
W.~Israel,
Nuovo Cim. B \textbf{44S10}, 1 (1966)
[erratum: Nuovo Cim. B \textbf{48}, 463 (1967)]. 

\bibitem{gibbhawk}
G.~W.~Gibbons and S.~W.~Hawking,
Phys. Rev. D \textbf{15}, 2752-2756 (1977). 

\bibitem{gibbhawkcosm}
G.~W.~Gibbons and S.~W.~Hawking,
Phys. Rev. D \textbf{15}, 2738-2751 (1977). 

\bibitem{Coleman:1977py}
S.~R.~Coleman,
Phys. Rev. D \textbf{15}, 2929-2936 (1977)
[erratum: Phys. Rev. D \textbf{16}, 1248 (1977)]. 

\bibitem{Callan:1977pt}
C.~G.~Callan, Jr. and S.~R.~Coleman,
Phys. Rev. D \textbf{16}, 1762-1768 (1977). 

\bibitem{Coleman:1980aw}
S.~R.~Coleman and F.~De Luccia,
Phys. Rev. D \textbf{21}, 3305 (1980). 

\bibitem{Bousso:2000xa}
R.~Bousso and J.~Polchinski,
JHEP \textbf{06}, 006 (2000) 
[arXiv:hep-th/0004134 [hep-th]].

\bibitem{Feng:2000if}
J.~L.~Feng, J.~March-Russell, S.~Sethi and F.~Wilczek,
Nucl. Phys. B \textbf{602}, 307-328 (2001) 
[arXiv:hep-th/0005276 [hep-th]].

\bibitem{Banks:2000fe}
T.~Banks,
Int. J. Mod. Phys. A \textbf{16}, 910-921 (2001)
[arXiv:hep-th/0007146 [hep-th]].

\bibitem{Banks:2001yp}
T.~Banks and W.~Fischler,
[arXiv:hep-th/0102077 [hep-th]].

\bibitem{Witten:2001kn}
E.~Witten,
[arXiv:hep-th/0106109 [hep-th]].

\bibitem{Goheer:2002vf}
N.~Goheer, M.~Kleban and L.~Susskind,
JHEP \textbf{07}, 056 (2003)
[arXiv:hep-th/0212209 [hep-th]].

\bibitem{Dvali:2017eba}
G.~Dvali, C.~Gomez and S.~Zell,
JCAP \textbf{06}, 028 (2017)
[arXiv:1701.08776 [hep-th]].

\bibitem{Obied:2018sgi}
G.~Obied, H.~Ooguri, L.~Spodyneiko and C.~Vafa,
[arXiv:1806.08362 [hep-th]].

\bibitem{Susskind:2021dfc}
L.~Susskind,
[arXiv:2109.01322 [hep-th]].

\bibitem{Coleman:1988tj}
S.~R.~Coleman,
Nucl. Phys. B \textbf{310}, 643-668 (1988). 

\bibitem{Coleman:1989ky}
S.~R.~Coleman and K.~M.~Lee,
Phys. Lett. B \textbf{221}, 242-249 (1989). 

\bibitem{Hawking:1978pog}
S.~W.~Hawking,
Nucl. Phys. B \textbf{144}, 349-362 (1978).

\bibitem{Banks:1984cw}
T.~Banks,
Nucl. Phys. B \textbf{249}, 332-360 (1985). 

\bibitem{Giddings:1988wv}
S.~B.~Giddings and A.~Strominger,
Nucl. Phys. B \textbf{321}, 481-508 (1989).

\bibitem{Fischler:1988ia}
W.~Fischler and L.~Susskind,
Phys. Lett. B \textbf{217}, 48-54 (1989)

\bibitem{Fischler:1989ka}
W.~Fischler, I.~R.~Klebanov, J.~Polchinski and L.~Susskind,
Nucl. Phys. B \textbf{327}, 157-177 (1989)

\bibitem{Polchinski:1989ae}
J.~Polchinski,
Nucl. Phys. B \textbf{325}, 619-630 (1989). 

\bibitem{Horava:2000tb}
P.~Ho\v rava and Dj.~Mini\'c,
Phys. Rev. Lett. \textbf{85}, 1610-1613 (2000)
[arXiv:hep-th/0001145 [hep-th]].

\bibitem{niven}
I. Niven, {\it Numbers: Rational and Irrational}, Mathematical 
Association of America: New Mathematical Library (June 1, 1961).

\bibitem{banksseiberg}
T.~Banks and N.~Seiberg,
Phys. Rev. D \textbf{83}, 084019 (2011). 

\bibitem{Garriga:2000cv}
J.~Garriga and A.~Vilenkin,
Phys. Rev. D \textbf{64}, 023517 (2001)
[arXiv:hep-th/0011262 [hep-th]].

\bibitem{Carroll:2004pn}
S.~M.~Carroll and J.~Chen,
[arXiv:hep-th/0410270 [hep-th]].

\bibitem{Bousso:2008hz}
R.~Bousso, B.~Freivogel and I.~S.~Yang,
Phys. Rev. D \textbf{79}, 063513 (2009)
[arXiv:0808.3770 [hep-th]].

\bibitem{DeSimone:2008if}
A.~De Simone, A.~H.~Guth, A.~D.~Linde, M.~Noorbala, M.~P.~Salem and A.~Vilenkin,
Phys. Rev. D \textbf{82}, 063520 (2010)
[arXiv:0808.3778 [hep-th]].

\bibitem{Susskind:2014rva}
L.~Susskind,
Fortsch. Phys. \textbf{64}, 24-43 (2016)
[arXiv:1403.5695 [hep-th]].

\end{thebibliography}
\end{document}